%% file: VAInpaint.tex
\documentclass{article}
\usepackage{spconf,amsmath,graphicx,hyperref}
\usepackage{amssymb}
\usepackage{booktabs}
\usepackage{subcaption}
\usepackage{tikz}    
\usepackage{pgfplots} 
\usepackage{float}
\pgfplotsset{compat=1.18}
\usepackage{verbatim}

\title{VAInpaint: Zero-Shot Video-Audio inpainting framework with LLMs-driven Module}
\name{Kam Man Wu \qquad Zeyue Tian \qquad Liya Ji \qquad Qifeng Chen$^{\dagger}$}
\address{The Hong Kong University of Science and Technology}

\begin{document}
%
\maketitle
\begin{abstract}
Video and audio inpainting for mixed audio-visual content has become a crucial task in multimedia editing recently. However, precisely removing an object and its corresponding audio from a video without affecting the rest of the scene remains a significant challenge. To address this, we propose VAInpaint, a novel pipeline that first utilizes a segmentation model to generate masks and guide a video inpainting model in removing objects. At the same time, an LLM then analyzes the scene globally, while a region-specific model provides localized descriptions. Both the overall and regional descriptions will be inputted into an LLM, which will refine the content and turn it into text queries for our text-driven audio separation model. Our audio separation model is fine-tuned on a customized dataset comprising segmented MUSIC instrument images and VGGSound backgrounds to enhance its generalization performance. Experiments show that our method achieves performance comparable to current benchmarks in both audio and video inpainting.

\end{abstract}
\section{Introduction}
\label{sec:intro}
In the era of multimodal content creation, the ability to separate and edit mixed audio-visual (AV) elements is a crucial part of video editing. Traditional methods in AV combination content are often limited to the handling method within the speech domain or struggle with tangled audio-visual signals in static scenes \cite{pu_2017}. Occasionally, when performing the visual-audio separation task in the outdoor scenes, issues with unwanted leftover audio or visual elements in the output can also lower perceived quality~\cite{tzinis_2020}. In this paper, we define a machine learning task as follows: Given a video with multiple objects, such as a scene where people are playing instruments, separate the instrument audio from the mixed soundtrack, and inpaint the video to remove the instrument and performer in a zero-shot manner. This paper addresses the challenge of isolating clean video and audio tracks from mixed content, using advanced segmentation, inpainting, and language-guided techniques.

Prior works have constructed important foundations for the components of audio-visual processing. However, they are usually limited by their focus on visual features or audio features individually, which have some potential alignments and connections to other modalities. In video inpainting, efforts have focused on temporal consistency and efficiency \cite{wang_2019}, with recent models like ProPainter \cite{zhou_2023_propaint} advancing transformer-based, mask-guided object removal. For segmentation, foundational models such as SAM2 \cite{ravi_2024} enable precise, prompt-driven object isolation. On the audio side, OmniSep, a query-based audio separation model \cite{cheng_2024}, supports omni-modal inputs, including text. Meanwhile, LLM-guided methods \cite{mo_2024} have emerged to align representations using human-like language. AV separation or inpainting pipelines are often built based on these techniques, using segmentation for initial source localization, as seen in The Sound-Of-Pixels \cite{zhao2018sound}, and SAVE \cite{nguyen_2025}. However, a key limitation is that these methods often lack robust LLM alignment for high-level reasoning, leading to imprecise segmentation or correspondence and an inability to fully remove object-related sounds. Luckily, a new object-to-text pipeline named Describe Anything \cite{lian_2025} provides valuable localized captions, which indicates the potential of the model for enabling precise audio separation and inpainting guidance.

As a result, to address the problem of the critical disconnection between high-precision visual inpainting and semantically accurate audio removal, we introduce VAInpaint, which unifies these elements for end-to-end AV inpainting. We first utilize SAM2 to segment out our needed object, then we generate a corresponding object mask to guide ProPainter in removing undesired visual elements. Next, we utilize an LLM to analyze the extracted frame for overall image understanding, while using Describe Anything to generate region-specific text descriptions using the previously generated masks. Combining the above outputs will lead to an LLM-refined text query for OmniSep as an input. For our training data, we developed a custom dataset generation pipeline: We first use SAM2 to segment instruments and performers from MUSIC videos~\cite{zhao2018sound}, creating masks at any resolution. VGGSound videos are then resized to match the resolution, and we mix the segmented MUSIC elements into VGGSound backgrounds~\cite{chen2020vggsound}. This results in a dataset of mixed audio-visual samples. To increase the separation effect, we fine-tune our audio separation model using our customized dataset. To conclude, our paper’s main contributions are:
\\
\vspace{-6mm}
\begin{itemize}
    
    \item An integrated hybrid workflow combining video inpainting, LLMs-based comprehensive and regional Scene Understanding with text, and text-query based audio separation.
    \vspace{-2mm}
    \item A new dataset blending MUSIC\cite{zhao2018sound} and VGGSound\cite{chen2020vggsound} with audio-visual content, supported by automated scripts to ensure scalability.
    \vspace{-2mm}
    \item A pipeline that assists LLMs in more effectively extracting and condensing content.
\end{itemize}

\section{METHODOLOGY}
\label{sec:format}
\subsection{Preliminaries}

OmniSep performs audio source separation in the spectrogram domain \cite{cheng_2024}. For a mixture spectrogram $m \in \mathbf{R}^{B \times 1 \times F \times T}$ and $N$ source embeddings $\{e_n\}_{n=1}^N \in \mathbf{R}^{B \times D}$ from multimodal inputs, the model predicts separation masks $\{p_n\}_{n=1}^N$ via:

\begin{equation}
\begin{aligned}
f_s &= \text{U-Net}(\log(m + \epsilon)) \\
f_e^n &= \sigma(\mathbf{W}_e e_n + \mathbf{b}_e) \\
p_n &= \sigma(\langle \mathbf{s} \odot f_e^n, f_s \rangle + b)
\end{aligned}
\end{equation}

where $\text{U-Net}$ is an encoder-decoder network, $\mathbf{W}_e \in \mathbf{R}^{C \times D}$, $\mathbf{b}_e \in \mathbf{R}^C$, $\mathbf{s} \in \mathbf{R}^C$, $b \in \mathbf{R}$ are learnable parameters, $\sigma$ is sigmoid activation, and $\langle \cdot, \cdot \rangle$ denotes inner product.

The training process minimizes the weighted BCE loss:
\begin{equation}
\mathcal{L} = \frac{1}{N} \sum_{n=1}^N \mathbf{E}_{f,t} \left[ \max(\log(1 + m), 10^{-3}) \cdot \text{BCE}(p_n, t_n) \right]
\end{equation}

with optional log-frequency warping. Separated sources are obtained as $\hat{s}_n = p_n \odot m$.

\subsection{LLMs-Generated Text Queries}
\begin{figure}[H]
    \centering
    \includegraphics[width=1.05\linewidth]{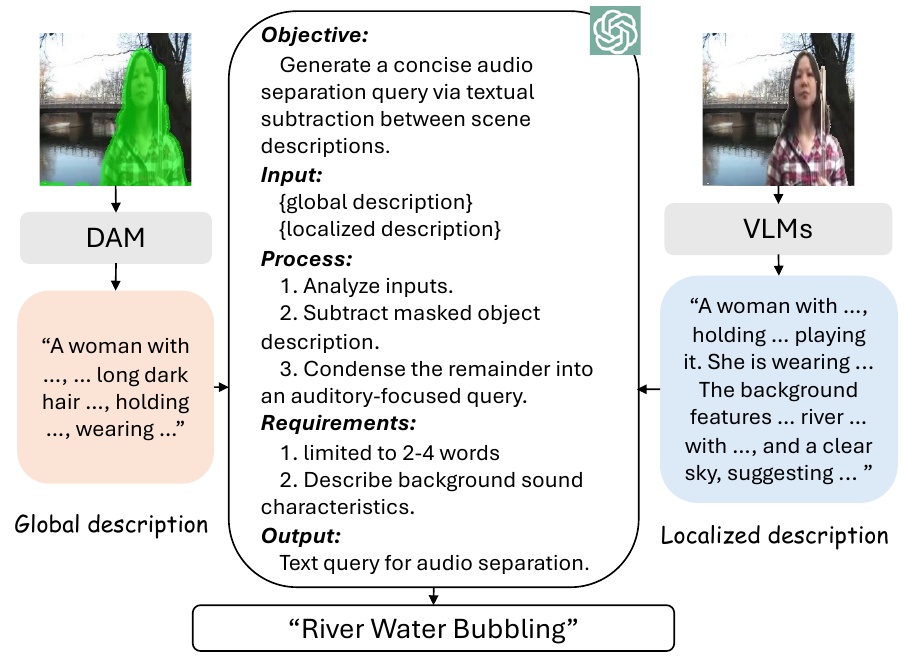}
    \vspace{-4mm}
    \caption{Pipeline of text query generation with LLMs.}
    \vspace{-4mm}
    \label{fig:text_query_generation}
\end{figure}

\begin{figure*}[t]
    \centering
    \includegraphics[width=0.9\linewidth]{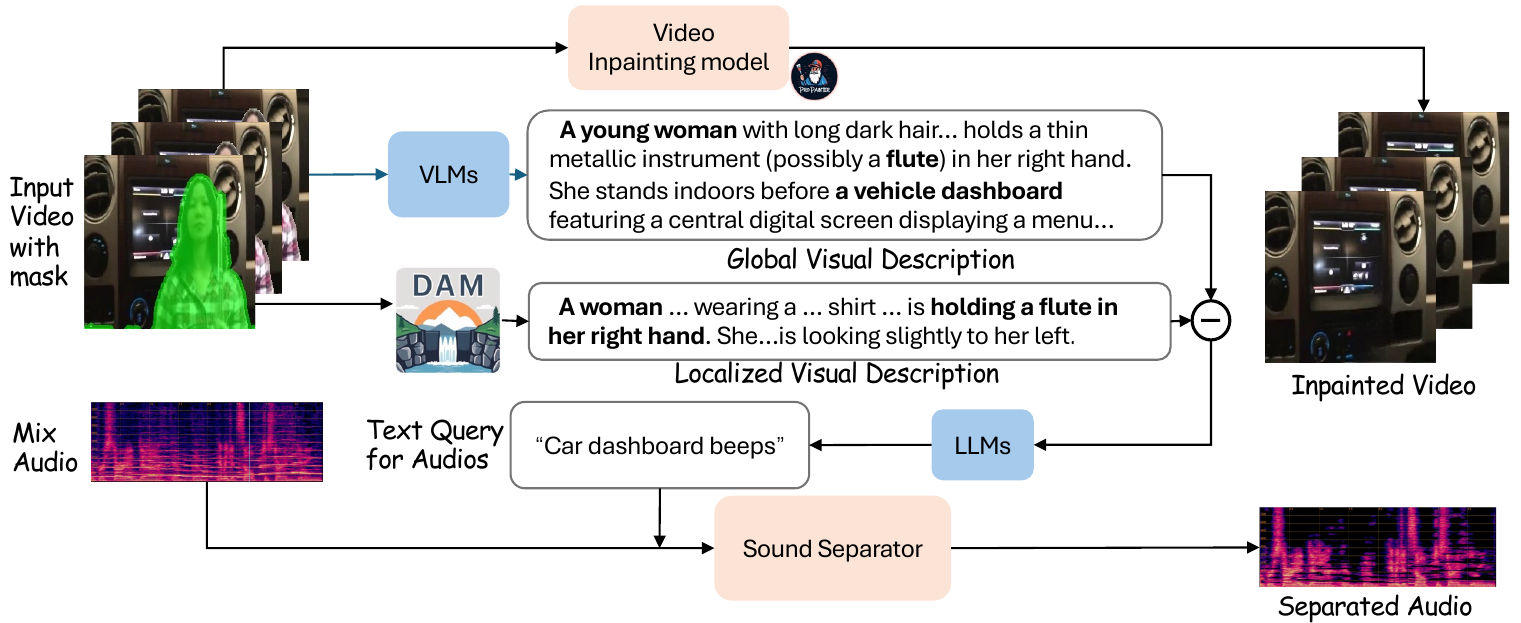}
    \vspace{-2mm}
    \caption{\textbf{Overview of our video-audio inpainting pipeline}. Taking the input video with the removed region and the corresponding mixed audio as the inputs, our method can correctly generate the inpainted video and separate the audio. We use the text modality to fill the domain gap between videos and audio. We found that LLMs can effectively transfer language from the domain of Visual-language models to that of text-audio models.}
    \label{fig:overall_workflow}
    \vspace{-2mm}
\end{figure*}

To convert visual-language descriptions into audio-language queries, we employ a two-step LLM process. First, a VLM analyzes an extracted frame for overall scene understanding: $ d_v = \text{VLM}(f) $, where $f$ is the extracted frame. Simultaneously, Describe Anything generates object-localized descriptions for the masked MUSIC region: $ d_a = \text{DescribeAnything}(f, m) $, with mask $ m $ from SAM2.
The LLM then performs textual subtraction and condensation:
\begin{equation}
    q = \text{LLM}(d_v - d_a)
\end{equation}
condensing the difference between two descriptions into an audio-focused query. As we can see from the figure \ref{fig:overall_workflow}, this text query guides OmniSep, which is our sound separator, for separation tasks. Our workflow transforms visual cues into auditory prompts, enabling the precise isolation of VGGSound audio \cite{chen2020vggsound}. For the details of the text query generation, we can refer to the content inside Figure \ref{fig:text_query_generation}. Inside, we describe how we utilize the Describe Anything Model (DAM) and VLM to generate regional descriptions and overall descriptions. Then we feed the descriptions into our used LLMs and generate a text query as requested.

\subsection{Model Fine-Tuning}

We fine-tune the OmniSep model \cite{cheng_2024} on our VAInpaint dataset using a supervised learning method. The model processes mixture spectrograms $m$ and multimodal embeddings $\{e_n\}_{n=1}^N$ (extracted using ImageBind \cite{girdhar_2023}) to predict separation masks $\{p_n\}_{n=1}^N$ through the architecture described in Section 2.1.

The model is optimized using the weighted binary cross-entropy loss:
\begin{equation}
\mathcal{L} = \frac{1}{N} \sum_{n=1}^N \mathbf{E}_{f,t} \left[ \max(\log(1 + m), 10^{-3}) \cdot \text{BCE}(p_n, g_n) \right]
\end{equation}
where $g_n$ denotes the ground-truth mask for source $n$.

\section{EXPERIMENTS}
\label{sec:experiments}

\subsection{Objective Metrics}
\label{subsec:objective_metrics}

We evaluate our model using the following objective metrics:

\subsubsection{Fréchet Distance (FD)}
FD quantifies the similarity between two multivariate Gaussian distributions fitted to feature embeddings \cite{kilgour_2018,hershey_2017}. The metric computes:
\begin{equation}
\mathcal{FD} = \|\mu_g - \mu_t\|^2 + \text{Tr}\left(\Sigma_g + \Sigma_t - 2(\Sigma_g\Sigma_t)^{1/2}\right),
\end{equation}
where $(\mu_g, \Sigma_g)$ and $(\mu_t, \Sigma_t)$ represent the mean and covariance of embeddings from generated and target features, respectively. Lower FD indicates better distribution alignment.

\input{fig_tex/fig3}
\subsubsection{Kullback-Leibler Divergence (KLD)}
KLD measures the distributional discrepancy between classifier outputs for separated and target audio using the binary formulation \cite{copet2023simple,haseeb2024gpt}:
\vspace{-1.5mm}
\begin{equation}
\mathcal{D}_{KL}(P\|Q) = \sum_i \left[ P_i \log\frac{P_i}{Q_i} + (1-P_i) \log\frac{1-P_i}{1-Q_i} \right],
\end{equation}
\vspace{-1mm}
where $P$ represents target class probabilities and $Q$ represents generated audio probabilities from sigmoid-activated outputs. Lower KLD values indicate better distribution matching.

\subsubsection{Scale-Invariant Signal-to-Distortion Ratio (SI-SDR)}
SI-SDR evaluates separation quality while remaining invariant to amplitude scaling \cite{le2019sdr}:
\begin{align}
\text{SI-SDR} &= 10 \log_{10} \frac{\|\alpha \mathbf{s}_t\|^2}{\|\mathbf{e}\|^2}, \\
\alpha &= \frac{\mathbf{s}_g^\top \mathbf{s}_t}{\|\mathbf{s}_t\|^2}, \quad 
\mathbf{e} = \mathbf{s}_g - \alpha \mathbf{s}_t,
\end{align}
where $\mathbf{s}_g$ is the separated signal and $\mathbf{s}_t$ is the target reference. Higher SI-SDR indicates superior separation performance.

\subsection{Experimental Results}
\begin{table}[H]
    \centering
    \scriptsize
    \setlength{\tabcolsep}{4pt}
    \scalebox{1.05}{
    \begin{tabular}{@{}lccrrrr@{}}
    \toprule
    Method  & Query Type & FD $\downarrow$ & KID $\downarrow$ & SI-SDR $\uparrow$ & SDR $\uparrow$ \\
    \midrule
    Sound-of-Pixels \cite{zhao2018sound} &  Visual & 95.32 & 11.16 & -24.48 & -1.66 \\
    iQuery \cite{chen2023iquery}&  Visual & 47.95 & 6.16 & -54.92 & -2.98 \\
    \midrule
    Ours (VGGSound Label)$^{*}$  & Text & 24.04 & 2.54 & 7.10 & 5.31 \\
    Ours (LLMs-Query)  & Text & 34.12 & 5.47 & -3.66 & 2.86 \\
    \textbf{Ours (LLMs-Query)$^{*}$}  & \textbf{Text} & \textbf{25.77} & \textbf{3.50} & \textbf{4.07} & \textbf{4.59} \\
    \bottomrule
    \end{tabular}}
    \vspace{-3mm}
    \caption{Quantitative results on audio separation. Our LLMs text query approach outperforms visual query baselines, with further improvements after finetuning the model on our designed dataset. $^{*}$ indicates the usage of the fine-tuned models.}
    \vspace{-4mm}
    \label{tab:objective_metrics1}
\end{table}
We evaluate our proposed OmniSep-LLMs method against the iQuery baseline \cite{chen2023iquery} and the Sound-of-Pixels baseline \cite{zhao2018sound} on our custom dataset. Our experimental results show that iQuery consistently fails to reproduce the correct audio pitch, producing outputs that are consistently lower than the ground truth. In contrast, our text-query pipeline accurately generates the pitch of separated audio sources. Meanwhile, the Sound-of-Pixels method frequently produces audio with significant data loss, which severely degrades both output audio quality and audience listening experience. Also, from the Table~\ref{tab:objective_metrics1} we can see that although the original OmniSep checkpoint substantially outperforms these visual query baselines, it remains limited by its reliance on precise text descriptions and struggles to distinguish between sources from the same category (e.g., differentiating between two instruments).

To improve generalization, we fine-tuned the model on our custom dataset. As shown in Table~\ref{tab:objective_metrics1}, finetuning results in substantial gains across all metrics. Qualitatively, the separation is significantly improved, with reduced artifacts and clearer output, as illustrated in Figure~\ref{fig:qualitativeResult}. Although directly using VGGSound labels is infeasible in real-life applications \cite{chen2020vggsound}, the proximity of our results to those obtained with ideal VGGSound labels indicates strong alignment between our LLMs-based visual-to-text pipeline and expert annotations. This demonstrates the effectiveness of our query generation approach. After finetuning, the model achieves more precise separation for both same-category and different-category audio mixtures.

\subsection{Ablation Study}
\begin{table}[H]
    \centering
    \scriptsize
    \setlength{\tabcolsep}{3pt}
    \scalebox{1.1}{
    \begin{tabular}{@{}l|c|c|rrrr@{}}
    \toprule
    Model & Type & FD $\downarrow$ & KLD $\downarrow$ & SI-SDR $\uparrow$ & SDR $\uparrow$ \\
    \midrule
    ChatGPT5  & General & 39.96 & 2.69 & -0.29 & 3.49 \\
    \textbf{ChatGPT5$^{*}$}  &\textbf{ General} & \textbf{19.02} & \textbf{0.91} & \textbf{6.74} & \textbf{5.29} \\
    \midrule
    Gemini-2.5-Flash & Multimodal & 84.92 & 3.62 & -10.81 & 3.62 \\
    Gemini-2.5-Flash$^{*}$ & Multimodal & 35.98 & 3.09 & 4.41 & 4.66 \\
    \midrule
    Grok4 & General & 44.62 & 2.71 & 0.63 & 4.39 \\
    Grok4$^{*}$ & General & 25.56 & 2.01 & 6.01 & 5.04 \\
    \midrule
    Qwen3-Max-Preview & Multimodal & 74.48 & 4.37 & -11.64 & 3.08 \\
    Qwen3-Max-Preview$^{*}$ & Multimodal & 35.83 & 3.25 & 2.66 & 4.14 \\
    \midrule
    DeepSeek-VL2-Small & Reasoning & 45.71 & 3.78 & -4.05 & 2.52 \\
    DeepSeek-VL2-Small$^{*}$ & Reasoning & 30.34 & 2.19 & 4.23 & 4.37 \\
    \bottomrule
    \end{tabular}}
    \caption{Objective metrics comparing audio separation performance, evaluated with synthetically generated text queries. $^{*}$ indicates the usage of a fine-tuned model.}
    \label{tab:objective_metrics2}
\end{table}

Our ablation study assesses the image-to-text capabilities of various large language models on our custom dataset, utilizing our audio separation pipeline. We assess models on image understanding, regional description, and query refinement through our pipeline to determine their effectiveness in generating user-wanted text queries. Table~\ref{tab:objective_metrics2} indicates that ChatGPT-5 and Grok-4 outperform other models in these tasks. While DeepSeek-VL2-Small is behind these leaders \cite{wu2024deepseek}, it demonstrates stronger reasoning ability than Gemini-2.5-Flash and Qwen3-Max-Preview. Crucially, our fine-tuned model consistently surpasses the original OmniSep checkpoint. The above results confirm that our LLM-driven approach significantly enhances audio separation performance, acting as an important component of our video-audio inpainting pipeline.

\subsection{Qualitative Results}
Inside the content in Figure \ref{fig:qualitativeResult}, Object 1 stands for the masked region; in the case of our qualitative results, Object 1 stands for the people and their instruments. Meanwhile, Object 2 represents the remaining elements within the overall scene. Based on the qualitative results from Figure \ref{fig:qualitativeResult}, our fine-tuned model's predicted audio spectrum closely matches the ground truth audio spectrum, demonstrating clearer separation with fewer artifacts and more alignment compared to the baseline (LLMs-Query method tested on the original checkpoint) and the iQuery prediction method. The qualitative result visually confirms the superior performance of our pipeline in audio separation.

\vspace{-6mm}
\section{CONCLUSION}
We present VAInpaint, a novel audio-visual inpainting pipeline integrating segmentation (SAM2), video inpainting (ProPainter), LLMs, and a query-based audio separation model (OmniSep). Our key innovation is an LLM-driven "textual subtraction" method that generates precise separation queries by contrasting global and regional image descriptions. Supported by a custom MUSIC-VGGSound dataset, our fine-tuned model demonstrates competitive performance against current benchmarks. Ablation studies confirm the superiority of high-performance LLMs and significant gains from the fine-tuning process on our custom dataset. Although the complexity of our pipeline presents challenges, this work still advances the field of automated audio-visual editing. Our future work will focus on dynamic multi-object scenes and extend the content of text queries for better Video-Audio inpainting control.
\vspace{-3mm}
\bibliographystyle{IEEEbib}
\bibliography{refs}

\end{document}

%% file: fig_tex/fig3.tex
\begin{figure*}[t]
    \centering
    \includegraphics[width=0.9\linewidth]{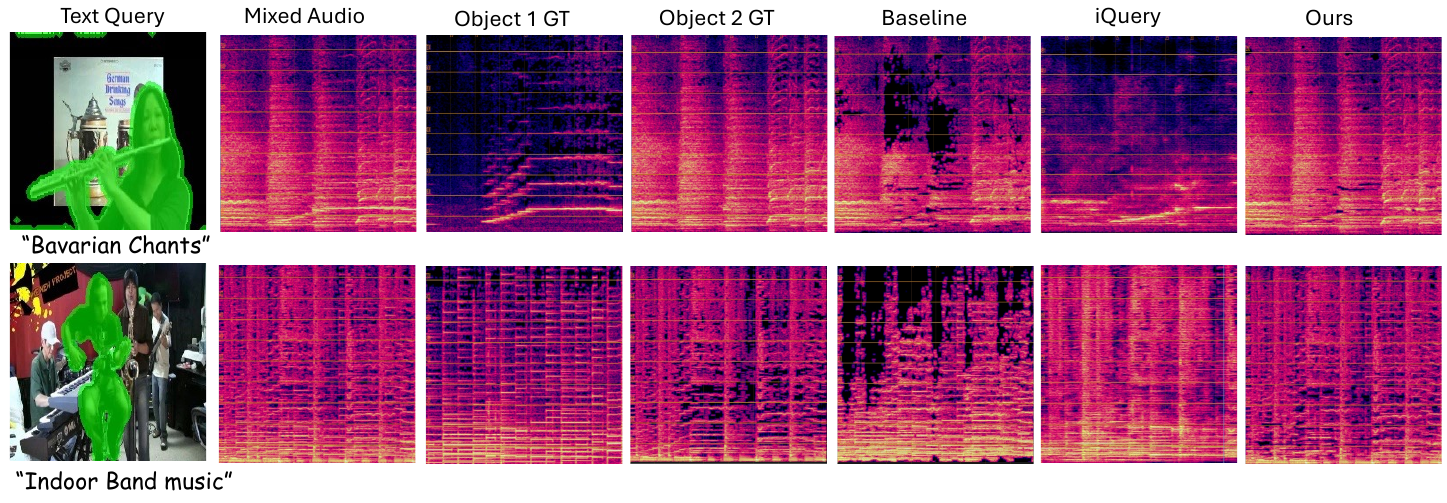}
    \vspace{-4mm}
    \caption{\textbf{Qualitative results for our methods.} The spectrum indicates that our 
    text-query pipeline (Original and Fine-tuned) produces cleaner source separation with fewer residual artifacts than the iQuery visual-query baseline.}
    \label{fig:qualitativeResult}
    \vspace{-4mm}
\end{figure*}